\documentclass[twocolumn,showpacs,aps,prd,floatfix,nofootinbib,superscriptaddress]{revtex4}
\usepackage{amsmath,xspace,array,longtable,dcolumn,hhline}
\usepackage{epsfig,graphicx,units,colordvi,amssymb,pifont,rotating}
\usepackage{relsize}
\def\babar{\mbox{\slshape B\kern-0.1em{\smaller A}\kern-0.1em
    B\kern-0.1em{\smaller A\kern-0.2em R}}}
\def\mmu         {\ensuremath{\mu}\xspace}
\def\mtau        {\ensuremath{\tau}\xspace}

\def\BBbar       {\ensuremath{B\overline B}\xspace}
\def\ee          {\ensuremath{e^+e^-}\xspace}
\def\epem        {\ensuremath{e^+e^-}\xspace}
\def\mumu        {\ensuremath{\mu^+\mu^-}\xspace}
\def\tautau      {\ensuremath{\tau^+\tau^-}\xspace}
\def\qqbar       {\ensuremath{q\overline q}\xspace}
\def\eemumu      {\ensuremath{\ee\rightarrow\mumu}\xspace}
\def\eetautau    {\ensuremath{\ee\rightarrow\tautau}\xspace}
\def\roots       {\ensuremath{\sqrt{s}}}
\def\rootsp      {\ensuremath{\sqrt{s^\prime}}}
\newcommand{\gev}{\ensuremath{\mathrm{\,Ge\kern -0.1em V}}\xspace}
\newcommand{\mev}{\ensuremath{\mathrm{\,Me\kern -0.1em V}}\xspace}
\def\pb          {\ensuremath{{\rm \,pb}}\xspace}

\def\nb          {\ensuremath{{\rm \,nb}}\xspace}
\def\mz          {\ensuremath{\mathrm{m_{Z}}}\xspace}
\def\kk          {\mbox{\tt KKMC}\xspace}
\def\tauola      {\mbox{\tt TAUOLA}\xspace}

\def\koralb      {\mbox{\tt KORALB}\xspace}

\def\sigmaone    {\ensuremath{\sigma^{\mbox {\tiny KORALB}}_{\mbox{\tiny BORN}}}}

\def\sigmatwo    {\ensuremath{\sigma^{\mbox {\tiny KK}}_{\mbox{\tiny BORN}}}}

\def\sigmathr    {\ensuremath{\sigma^{\mbox {\tiny KORALB}}}}

\def\sigmafor    {\ensuremath{\sigma^{\mbox {\tiny KORALB}}_{\mbox{\tiny NO~VP}}}}

\def\sigmafiv    {\ensuremath{\sigma^{\mbox {\tiny KK}}_{\mbox{\tiny NO~BREM}}}}

\def\sigmasix    {\ensuremath{\sigma^{\mbox {\tiny KK}}_{\mbox{\tiny NO~VP}}}}

\def\sigmakk     {\ensuremath{\sigma^{\mbox {\tiny KK}}}}

\def\sigmakknoint{\ensuremath{\sigma^{\mbox {\tiny KK}}_{\mbox{\tiny NO~INT}}}}

\begin{document}

\title{
{\large \bf \boldmath
Tau and muon pair production cross sections in\\
electron-positron annihilations at \roots\ = 10.58 \gev}
}

\author{Swagato Banerjee}
\affiliation{Dept. Physics and Astronomy, University of Victoria, Victoria, British Columbia, Canada}
\author{Bolek Pietrzyk}
\affiliation{Laboratoire d'Annecy-le-Vieux de Physique des
Particules LAPP, IN2P3/CNRS, Universit\'e de Savoie, F-74019
Annecy-le-Vieux cedex, France}
\author{J.~Michael Roney}
\affiliation{Dept. Physics and Astronomy, University of Victoria, Victoria, British Columbia, Canada}
\author{Zbigniew Was}
\affiliation{Institute of Nuclear Physics, P.A.N. ul. Radzikowskiego 152 PL-31342 Krak\'{o}w, Poland}

\date{\today}

\begin{abstract}
The calculational precision of $\ee\to\tautau$ and $\ee\to\mumu$ production cross sections in
electron-positron annihilations at \roots\ = 10.58 \gev
is studied for the \kk Monte Carlo simulation program,
modified to include contributions from recent implementation of the hadronic part of vacuum polarization.
We determine $\sigma (\ee\to\tautau) = (0.919 \pm 0.003) \nb$ and
$\sigma(\ee\to\mumu) = (1.147 \pm 0.005) \nb$, where the error
represents the precision of the calculation.
\end{abstract}


\maketitle

\setcounter{footnote}{0}

\section{Introduction}
\label{Introduction}

At the present generation B-Factories lepton pair production,  $\ee\to\tautau$ and $\ee\to\mumu$,
occurs at approximately the same rate as the $\ee\to\BBbar$ process.
Given the high statistics of the available data,
precise knowledge of the \mtau-pair and \mmu-pair production cross sections in \epem annihilations
at a center-of-mass energy \roots\ = 10.58 \gev is necessary
for a number of  precision measurements,
particularly for measurements of the branching fractions of $\tau$ decays
and luminosity determinations using the counting of $\ee\to\mumu$ events.
The Belle experiment currently uses a $\tau$-pair cross section of 0.89~nb and quotes
a 1.4\% error on the number of $\tau$-pairs produced for a given amount of integrated
luminosity where the 1.4\% is intended to account mainly for the luminosity measurement
 error~\cite{Hayasaka:2007vc,Yusa:2006qq}.
The \babar\ experiment also uses a 0.89~nb cross section but assigns a 2\% error to this figure based
on a simple comparison of the cross sections from the  \kk~\cite{Jadach:1999vf} and \koralb~\cite{Jadach:1994ps}
Monte Carlo simulation programs~\cite{Aubert:2005wa}.
Although the 2\% figure is adequate  in searches for rare or forbidden processes, such a large
error currently limits the precision of $\tau$ branching fraction measurements at the 
B-factories\footnote{An alternative approach that is insensitive to \mtau-pair production cross section
is to measure the \mtau-branching ratios relative to the
 $\tau$ electronic branching fraction which is measured to $\sim 0.3\%$ accuracy, 
as has been done in a recent publication~\cite{Epifanov:2007rf}.
However, in such an approach, the corresponding Monte Carlo generators, e.g. \tauola~\cite{Jezabek:1991qp}, 
would need to be tested at the precision of 0.1\% level in the presence of experimental cuts,
and such a study is beyond the scope of the present paper.}.

In principle, the cross sections determined with \kk are the most precise calculations
currently available, but the cross sections and their precisions as determined with this
code at energies far below the $Z$-pole have not been reported.
Only tests for $\sqrt{s}>40\gev$ have been available in the
 literature~\cite{Jadach:1999vf,Jadach:2000ir,Ward:2002qq} and the code
cautions users of its potential lack of reliability for energies below $\sqrt{s}=40~\gev$
because of insufficient study in that regime.

This unsatisfactory situation should be compared with predictions of \kk for high energies,
where a precision at the 0.1-0.4\% level was achieved~\cite{Jadach:2000ir,Ward:2002qq}.
In this paper we report on straightforward studies of the $\mu$ and $\tau$ pair production cross-section
calculations of  \kk and  \koralb at \roots\ = 10.58 \gev with a goal of providing, for the
first time, a realistic estimate of the precision of the cross-section calculation
at energies relevant to the B-factories relying on, but not repeating, the detailed
high energy investigations.
The cross sections are calculated using code that includes
QED and electroweak radiative corrections
to different orders in the fine structure constant $\alpha$
and using different approximations for vacuum polarization.
\koralb calculates the matrix element to order $\alpha$ including interference between
initial (ISR) and final state radiation (FSR).
\kk calculates the cross section to order $\alpha^2 \log(s/m_e^2)$
and includes exponentiation and ISR-FSR interference.
Various implementations of the vacuum polarization can be used in
both \koralb and \kk programs.

 The results presented here rely on the extensive studies~\cite{Jadach:2000ir,Ward:2002qq} used
to determine the precision of the calculations in the  \kk as well as KORALZ~\cite{Jadach:1991ws} software.
The conclusion of those studies is that cross sections calculated with \kk  at the $Z$-pole and
LEP~II energies ($\roots \sim 200~\gev$) are at the precision level of 0.2\%. Fortunately,
the physics environment at the higher energies where the detailed tests have been made
 is far more complex than at the $\Upsilon(4S)$. For example, unlike at the   $\Upsilon(4S)$,
 the higher energy tests had to address issues related to the radiative return to the $Z$-pole, 
 including effects related to initial and final state radiation bremsstrahlung interference; scattering-angle
 dependencies, of particular interest because of the importance of forward-backward asymmetry measurements at the higher energies;
  and high-energy electroweak and QCD corrections. 
 Therefore it is only necessary to verify that the technical precision determined
 in detail at the higher energies for a wide class of effects can be reliably extended
 to the lower energies of the B-factories.
 This has been accomplished using relatively simple tests.
 In this way we achieve the goal of providing a realistic cross-section error without having to repeat the
 previous high energy studies.
 However, because of the different energy scales, the high energy studies of several effects, such as
 the vacuum polarization, are not reliably extended to the $\Upsilon(4S)$, and these
 have been examined separately.

As it turns out, our current knowledge of the lepton pair production cross sections is
in fact partially limited by the treatment of vacuum polarization in this calculation.
In this paper, we study their contributions using different parameterizations
of the hadronic part of the vacuum polarization,
as well as our knowledge of the ratio $R = (\epem\to\qqbar)/(\epem\to\mumu)$.
Other  sources of theoretical uncertainty
arise from the implementation of initial and final state radiation
including their interference and electroweak corrections,
effects from virtual pair-productions of additional fermions, and the impact of
low-energy resonances on the cross section.

At the level of precision targeted in this work, cross sections
for \mtau-pair and \mmu-pair processes are adequately studied with 10$^6$ generated events.
Because of the larger contribution from radiative returns to very low energy
in \mmu-pair events, cross sections are also studied with an angular
acceptance in the center-of-mass for both muons of  $[22^\circ, 158^\circ]$
and with effective center-of-mass energy, $\sqrt{s^{\prime}}$,  greater than 10\% of 10.58 GeV
which corresponds more closely to the actual experimental acceptance.

Our paper is organized as follows. In Section~\ref{VacPol} we discuss radiative corrections 
to the total cross section due to hadronic vacuum polarization. We have found that changes
to the public versions of the two programs are necessary in order to achieve the goal of
obtaining a  reduced systematic error. Section~\ref{ISRFSR} is devoted to verifying
 that the high energy estimates of the systematic error of \kk photonic (and acceptance dependent) corrections 
 for the total cross section remain valid at our energies as well.  
 Sections~\ref{Interfer} and ~\ref{PairProduction}
 discuss the impact of interference and pair production corrections, respectively. Section~\ref{Resonances}
describes how we address the effects of vector resonances. Section~\ref{Ratio} discusses the
error on the ratio of the \tautau and \mumu cross sections as calculated by \kk.
The estimations of the systematic error of theoretical predictions for the $\tau$-pair and
 $\mu$-pair cross sections are summarized in Section~\ref{Summary}.
 
\section{Treatment of vacuum polarization}
\label{VacPol}

The particular vacuum polarization implementation is controlled in the \kk code
using the IHVP flag following the DIZET implementation of the calculation~\cite{Bardin:1989tq},
with improvements summarized in section 4.1.3 of Ref.~\cite{Bardin:1999yd} and section 4.132 of Ref.~\cite{Kobel:2000aw}.
The default modeling of vacuum polarization in \kk corresponding to IHVP flag = 1~\cite{Eidelman:1995ny}, 
suggested in Ref.~\cite{Bardin:1999yd} was not optimized for the low energy applications.
In the present implementation, the hadronic part of the vacuum polarization is studied using
the experimental knowledge of the ratio $R = \sigma(\epem\to\qqbar)/\sigma(\epem\to\mumu)$ 
relevant for a precise determination of $\alpha(\mz)$.
The hadronic part of the vacuum polarization is not calculated in this option
in DIZET for center-of-mass energies less than 40~\gev.
Thus, at \roots = 10.58~\gev, this option gives a $\sim 3\%$ lower fermion-pair cross section
than those obtained from other DIZET options that include the hadronic component of the
vacuum polarization.
Other available options in DIZET (IHVP flag = 2~\cite{Jegerlehner:1991ed} 
and = 3~\cite{Burkhardt:1989ky}) use experimental knowledge of the ratio 
$R = (\epem\to\qqbar)/(\epem\to\mumu)$ measured more than 15 years ago. 
Improvements in the measurements of the ratio $R$ 
from the Crystal Ball experiment~\cite{CrystalBall} 
in \epem annihilations at $\roots$ between 5.0 and 7.4~\gev, 
and more recent measurements by the BES experiment~\cite{Bai:2001ct}
at $\roots$ between 2 and 5~\gev, are not included for these options.
Improvements on the calculation of the  hadronic part of the vacuum polarization
at $\roots = \mz$ obtained by including these new measurements
of $R$ have been documented recently~\cite{BuPie}.
Following these developments, we have introduced a new option in DIZET 
implemented with IHVP flag = 4 to calculate the contribution of vacuum polarization 
using the routine ``REPI''~\cite{Burkhardt:2005se}, where  these improvements 
on the measurement of $R$ have been incorporated.

We evalute the uncertainties associated with using IHVP flag = 4.
The routine REPI uses a simple parametrization of the hadronic contribution
to the vacuum polarization in the $t$-channel
and can be used safely in the $s$-channel
outside the energy regime dominated by individual resonances. 
Special care must be taken to account for the large positive and negative 
fluctuations in the contribution to the hadronic part of the vacuum polarization
in the $s$-channel across the resonance as illustrated in FIG.~1 of
Ref.~\cite{Karlen:2001hw}. We have calculated the differences
between the $s$-channel and $t$-channel contribution of the hadronic part
of the vacuum polarization to the \mtau and \mmu-pair production cross section,
weighted by the distribution of the invariant mass of the lepton-pair system (\rootsp)
in bins of 10~\mev. Below and above the resonances, these differences are 
positive and negative, respectively, and their effects tend to cancel when integrated
over \rootsp. At the $\Upsilon(4S)$ these differences cancel partially, and 
the size of this cancelation is sensitive to the variation of the
central value of the center-of-mass energy distribution of
colliding beams for different periods of data taking as well as
the collider beam energy spread. We assume a beam energy spread
 of 4.6~\mev(RMS),  which is characteristic of the center-of-mass
energy spread at \babar\ experiment at the PEP~II B-factory~\cite{Aubert:2004pwa}.
 Based on the results of these calculations
 we assign an uncertainty of 0.18\% on the \mtau-pair cross section 
 and 0.22\% for the \mmu-pair cross section
associated with the implementation of the vacuum polarization in \kk
with IHVP flag = 4 in DIZET.
These uncertainties are also valid for the Belle experiment,
since the center-of-mass energy spread of the KEK B-factory 
is similar to that of the PEP~II B-factory~\cite{KEKB}.

In order to check the technical precision of this new implementation of
vacuum polarization, we have also calculated the cross section with \kk
using different IHVP options in DIZET as well as with corresponding
calculations using \koralb  with this new IHVP flag = 4 option implemented.
The cross sections for $\eetautau$ and $\eemumu$
 calculated with \kk are summarized in Table~\ref{tab:1} along with their
Monte Carlo statistical errors, which are at the level of 0.02\% for \mtau-pairs
and 0.03\% for the \mmu-pairs.
The cross sections for \mmu-pairs with the cuts mentioned in the previous section
are denoted with ``${acc}$'' and ``${cuts}$'' subscripts, respectively,
for the cases when only the angular acceptance cuts are applied
and when both the acceptance and the $\rootsp > 0.1 \roots$ cuts are applied.
For $\tau$-pair production, we report the total cross section.

\begin{table*}[!ht]
\caption{$\tau$-pair and $\mu$-pair cross sections with different IHVP flags using \kk.
Note: IHVP=1 does not include the hadronic part of the vacuum polarization.}
\label{tab:1}
\begin{ruledtabular}
\begin{tabular}{l | c | c | c | c} 
IHVP & $\sigma(\tautau)$(\pb) & $\sigma(\mumu)$(\pb) & $\sigma_{acc}(\mumu)$(\pb) & $\sigma_{cuts}(\mumu)$(\pb) \\\hline
 1   & 891.45 $\pm$ 0.20      &  1117.81 $\pm$ 0.35  &  825.90 $\pm$  0.26 & 810.75 $\pm$ 0.25  \\
 2   & 919.85 $\pm$ 0.21      &  1149.01 $\pm$ 0.37  &  851.47 $\pm$  0.27 & 836.17 $\pm$ 0.27  \\
 3   & 920.17 $\pm$ 0.21      &  1148.56 $\pm$ 0.37  &  851.69 $\pm$  0.27 & 836.47 $\pm$ 0.27  \\
 4   & 918.66 $\pm$ 0.21      &  1146.60 $\pm$ 0.36  &  850.31 $\pm$  0.27 & 835.15 $\pm$ 0.27  \\
\end{tabular}
\end{ruledtabular}
\end{table*}

\begin{figure}[!hbtp]
\resizebox{.96\columnwidth}{.3\textheight}{%
\includegraphics{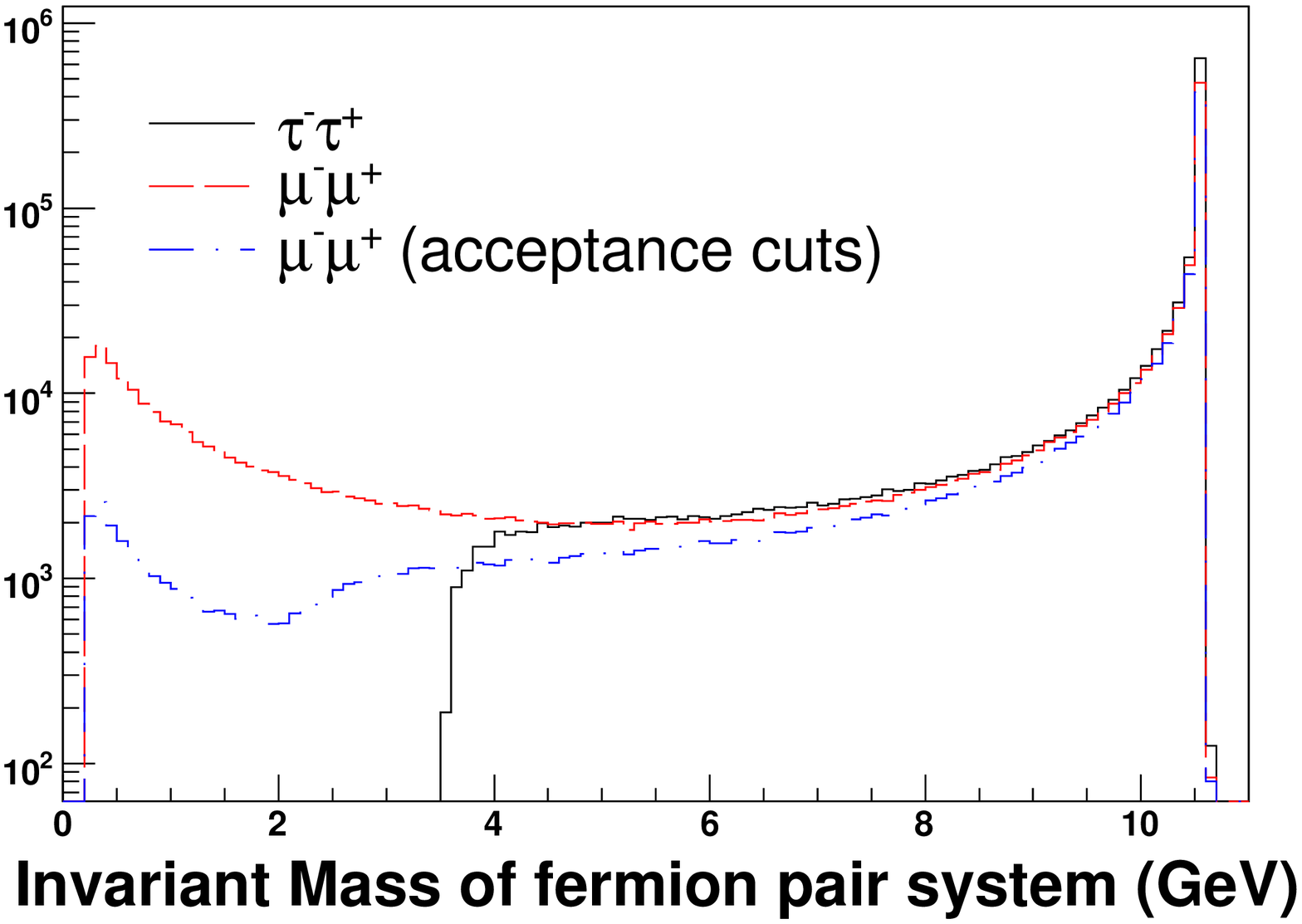}}
\caption{Invariant Mass of \tautau and \mumu pairs ($\sqrt{s^{\prime}}$) with IHVP flag = 4.}
\label{fig:1}
\end{figure}

The \mtau-pair cross sections from IHVP flags = 2, 3 and 4 agree
to within 0.16\%. The \mmu-pair cross sections agree within 0.16\%
and 0.21\% for the no-cuts case, $\sigma(\mumu)$, and all cuts
case, $\sigma_{cuts}(\mumu)$, respectively. As is evident from the
distribution of the lepton-pair invariant mass, \rootsp,
presented in FIG.~\ref{fig:1}, the \mmu-pair cross section has a
larger contribution from the vacuum polarization at low \rootsp.
Differences in the cross sections are expected because different measurements
of $R$ are used in the calculations of the vacuum polarization in
Refs.~\cite{Jegerlehner:1991ed,Burkhardt:1989ky,Burkhardt:2005se},
as described above. Therefore we do not assign any additional uncertainty
resulting from this cross check, when IHVP flag = 4 is used as the default option.

\begin{table*}[!ht]
\caption{Code configurations. Note: for technical reasons, the \kk code can only be
 run with final state radiation on.}
\label{tab:2}
\begin{ruledtabular}
\begin{tabular}{l | c | c | c } 
            & Software  & bremsstrahlung      & electroweak and vacuum        \\
            & Package   & configuration        & polarization corrections      \\ \hline
\sigmaone   & \koralb   & ISR off, FSR off     & off                    \\
\sigmatwo   & \kk       & ISR off, FSR on      & off                    \\
\sigmathr   & \koralb   & on with interference & on                      \\
\sigmafor   & \koralb   & on with interference & off                    \\
\sigmafiv   & \kk       & ISR off, FSR on      & on                     \\
\sigmasix   & \kk       & on with interference & off                    \\
\sigmakk   & \kk        & on with interference & on                     \\
\sigmakknoint   & \kk        & ISR/FSR on but interference off & on                     \\
\end{tabular}
\end{ruledtabular}
\end{table*}

As an additional cross check on the introduction  of the vacuum polarization into the code,
we have studied the technical robustness achieved in the implementation of 
IHVP flag = 4 by comparing the cross sections calculated with \koralb and \kk
programs using the different configurations given in Table~\ref{tab:2}.
Turning off the bremsstrahlung radiative corrections allows us to isolate
the vacuum polarization corrections. In the case of \koralb, 
the vacuum polarization increases the Born-level cross section (\sigmaone)
by $2{\cal R}e(\Pi_{\gamma\gamma}(s^{\prime}=s))$,
where $\Pi_{\gamma\gamma}$ is the photon self-energy function for the vacuum polarization correction.
On the other hand, for \kk, the vacuum polarization factor is
${|1-\Pi_{\gamma\gamma}(s^\prime)|}^{-2}$, where $s^{\prime}$ is only determined from ISR,
so that for the case of FSR alone, $s=s^{\prime}$.
Therefore, we expect
$${\left(\sigmathr-\sigmafor\right)\over{\sigmaone}} \approx 2 \left(1-\sqrt{{\sigmatwo}\over{\sigmafiv}}\right)\cdot$$
This relation between \kk and \koralb is found to be valid to better than the statistical precision of the
Monte Carlo calculations for both \mtau-pairs and \mmu-pairs 
with and without additional cuts on acceptance and \rootsp.

These results are consistent with our expectations. 
However, in general when ISR is turned on in \koralb  $s^{\prime}=s$ is always
used for the vacuum polarization calculation, whereas, in \kk $s^{\prime}$, more correctly,
varies from event to event as expected in the presence of ISR.
In the absence of ISR, the above relation comparing \kk and \koralb cross sections is exactly valid,
whereas in the presence of ISR, the difference in treatment of this effect between the two programs
is small enough that the relation still provides a useful cross check.

\section{Initial and final state bremsstrahlung}
\label{ISRFSR}

Because bremsstrahlung is calculated to different orders in \kk and \koralb,
comparisons of the cross sections calculated using the two software packages
help  study properties  of the corrections,
and as a consequence, establish the precision for the \kk\ cross sections.
To that end,  we  compare results from Born-level calculations (\kk or \koralb),
first order radiative effects (\koralb)
and second order calculations with exponentiation (\kk).
For these studies, the electroweak and vacuum polarization corrections are initially switched off.

As a technical benchmark, we initially compare the Born-level
\eemumu\ cross sections calculated at $\roots = 10.58~\gev$ using
\kk and \koralb  when both electroweak and radiative corrections are switched off.
For technical reasons~\cite{Was:2006my}, the final state radiation in \kk is not swiched off and
therefore it is expected that the ratio $\sigmatwo/\sigmaone$ is equal to $1 + ({3\over4})({\alpha\over\pi})$
when no experimental cuts are applied. To better than 0.01\%, we find this to be the case.

Having established that the Born-level cross sections are in good agreement,
we proceed to use the \koralb cross sections with ISR-FSR switched on  and
with electroweak and vacuum polarization corrections switched off (\sigmafor).
The quantity $(1-\sigmafor/\sigmaone)$ is a measure of the contribution of first order bremsstrahlung to the
cross section.  The magnitude of the bremsstralung effects at the second and third orders in the exponential expansion
can be estimated as $(1-\sigmafor/\sigmaone)^2/2$ and $(1-\sigmafor/\sigmaone)^3/6$, respectively.
For $\sigma_{cuts}(\mumu)$, $\sigmafor/\sigmaone=1.1126$ whereas for $\sigma(\tautau)$,
$\sigmafor/\sigmaone=1.1054$. Taking 11\% as the size of first-order correction from bremsstrahlung
the second order in the exponential expansion series is $\sim 0.11^2/2=0.0061$.
Therefore, since \kk includes the second order, we expect a difference between the \kk and \koralb cross sections
to be {\footnotesize $\stackrel{<}{\sim}$} $1\%$. This 1\% estimate is consistent with the value
 of 1.0072 for $\sigmasix/\sigmafor$ for  $\sigma_{cuts}(\mumu)$,
and somewhat larger than the value of 1.0012 for  $\sigma(\tautau)$, which also receives contributions from other effects,
e.g. mass corrections. The next order of the leading
log terms, which is not fully controlled in \kk, can be estimated to be $\sim 0.11^3/6=0.0002$.
These tests confirm that \kk and \koralb are yielding the expected relative
 cross section behaviour near $\roots=10.58~\gev$.
We can estimate the relative error associated with the treatment of ISR and FSR
 to be equal to the size of the last term fully controlled in
the  approximation  implemented in \kk: $\alpha^2\log(s/m_e^2)=0.0011$.
This next-to-leading-log term is larger than the third order
 leading log term, which is expected to be of order of $\sim 0.11^3/6$, by about
a factor of five. 
The above estimates are rather na\"{\i}ve as they assume an exponentiation 
pattern for all QED effects. Nevertheless, in our case the resulting relations hold
rather well, pointing to a simple pattern of QED effects, 
and at the same time to the technical correctness of our calculations.
This is to be compared with the 0.2\% uncertainty assigned for the \kk
treatment of bremsstrahlung at LEP II energies~\cite{Ward:2002qq}, 
where further complications arise due to significant contributions from the radiative return to $Z$.

Final state bremsstrahlung has mass-term dependencies that potentially are more significant
for $\tau$-pair production. These would be expected to contribute a maximum 
of $\frac{\alpha}{\pi}\frac{4m_\tau^2}{s}=0.03\%$ to the relative systematic error
and therefore are negligible for this level of study and not considered further.

As these tests assume \koralb is correctly calculating 
 the matrix element to order $\alpha$, it is useful to ensure that its behaviour
associated with the technical parameter {\tt XK0}, the minimal energy for the real
bremsstrahlung photon to be explicitly generated, is insensitive to 
the actual choice of this parameter~\cite{Berends:1982ie}. To that end, we have 
 verified that reducing {\tt XPAR{11}=XK0} from its default value of 0.01 by a factor of two changes total
\koralb cross section by considerably less than 0.1\%.

Finally, as a technical cross check on the final results,
we also compare the \kk cross-section calculation with all
radiative corrections turned on, \sigmakk, to a cross-section calculation
assuming na\"{\i}ve factorization of the bremsstrahlung and
non-bremsstrahlung corrections: $\sigmakk = \sigmasix \times
(\sigmafiv/\sigmatwo)$.
For \mmu-pairs with $\rootsp > 0.9 \roots$,
which is in the regime of very small radiation,
we find that this factorization holds to better than the statistical precision
of the Monte Carlo calculations.

These studies verify that the 0.1-0.2\% error  tag  assigned for photonic 
corrections in Refs.~\cite{Jadach:2000ir,Ward:2002qq} for higher energies
is also valid at $\roots=10.58\gev$. We adopt the higher, 0.20\%, value
 as the uncertainty on the implementation of ISR/FSR bremsstrahlung at the $\Upsilon(4S)$
and thereby avoid having to address the dependence on selection cuts.

\section{Interference effects}
\label{Interfer}

Two kind of interference effects are considered:
the electroweak interference between $\gamma$ and $Z^\star$ propagators,
and the QED interference between
intial and final state radiation arising from $\gamma\gamma$ box diagrams.
The contribution from the $\gamma$-$Z^\star$ interference
is smaller than the Monte Carlo statistical precision.
Although the QED interference results in a forward-backward asymmetry
of a few percent,  the  contribution to the total cross section is considerably smaller.
The magnitude of the contribution of interference to the total cross section
can be estimated from the ratio of the cross section with all corrections on, $\sigmakk$,
to the cross section calculated with all corrections on but with interference switched off, $\sigmakknoint$.
For both \mmu-pairs and \mtau-pairs, $\sigmakk/\sigmakknoint = 1.0004$ and is relatively
insensitive to whether or not geometrically symmetric cuts are applied.
We conclude that the largest contribution to
uncertainties from these interference effects are of the order of 0.04\%.

\section{Pair-production and Vertex Correction Uncertainty}
\label{PairProduction}

The default calculations in \kk do not include the effects of the
emission of an extra fermion pair accompanying the main process
and the virtual fermion loop vertex correction. As these two effects
are of opposite sign and comparable magnitude they will largely cancel 
each other and the error introduced by neglecting them both in 
 the calculation is small. This large cancelation is a well-known feature of
QED and has been numerically verified by studies at higher energies~\cite{Kobel:2000aw}.

To estimate the error this introduces on the total  cross section,
 we  activate the appropriate second order contribution to the
 vertex in \kk as explained in Section 4.54 of Ref.~\cite{Kobel:2000aw},
 and supplement the simulation with the additional
sample of the four fermion final states, e.g. with the help of
KORALW~\cite{Jadach:2001mp}. The effect of the virtual fermion loop
vertex correction  at the $\Upsilon(4S)$, using
appropriate options in \kk, is  0.3\%. However, because
emission of real pairs largely cancels contribution from this vertex correction,
we can conservatively assign half of this 0.3\% as the error 
associated with ignoring both the pair-production and vertex virtual
fermion loop corrections. We therefore assign a contribution of 0.15\% to the
uncertainty on the cross sections.

\section{Impact of Resonances}
\label{Resonances}
All of the studies so far discussed have assumed 
$\ee\to\tautau$ and $\ee\to\mumu$ production to proceed via   
 $\gamma$ and $Z^\star$ propagators. However, at $\roots=10.58~\gev$ it is
also possible for a non-negligible fraction of the final state lepton-pair
yield to arise from the production and decay of intermediate
vector meson resonances, $J/\Psi, \Psi(2S), \Upsilon(3S), \Upsilon(4S)$ etc,
when accompanied by hard ISR: e.g. $\ee\to J/\Psi \gamma \to \mumu\gamma$.
We estimate the potential size of these contributions and add a systematic
error that accounts for the fact that they are not included in the \kk
program.


We calculate the cross section associated with the radiative return to each resonance
below the $\Upsilon(4S)$ using the narrow width approximation and a radiator function calculated to
order $\alpha^2$~\cite{Kuraev:1985hb,Benayoun:1999hm}. Updated resonance parameters
and leptonic branching ratios of reference ~\cite{Yao:2006px} are employed. 
In the case of  the $J/\Psi$, we can cross check the calculation against the \babar\ experiment's measurement of 
$\ee\to J/\Psi \gamma \to \mumu\gamma$~\cite{Aubert:2003sv} and find they agree to 7\%.
Table~\ref{tab:3} lists the
vector meson resonances that give a small contribution to the cross section.
Of the other vector meson resonances of potential interest ($\Psi(4415)$, $\Psi(4160)$, $\Psi(4039)$,$\Psi(3770)$,
$\rho(2150)$, $\rho(1900)$,  $\rho(1700)$, $\phi(1680)$, $\omega(1650)$,
 $\pi_1(1600)$,    $\rho(1450)$,     $\omega(1420)$,  $\pi_1(1400)$, $\phi(1020)$, $\omega(782)$ and $\rho(770)$)
 none contribute to the $\sigma_{cuts}(\mumu)$ or $\sigma(\tautau)$ cross sections at the levels of interest
 because they are either below the $\mu$-pair $\rootsp$ cut or $\tau$-pair threshold,
 or because they have a negligible leptonic branching ratio. For the $\Upsilon(4S)$, we use the
 measured cross section of 1.101~nb at the peak of the resonance~\cite{Aubert:2004pwa}. 
 The calculation of the contribution of the resonances on $\sigma_{cuts}(\mumu)$ employ
 the \kk Monte Carlo simulation to estimate the impact of the 
cuts\footnote{Note that the method employed to estimate the contributions to the resonances has
been improved compared to that used in the first version of this paper (arXiv:0706.3235v1 [hep-ph])
and results in a slightly larger determination of the total uncertainty.}.

\begin{table*}[!ht]
\caption{Estimates of intermediate vector resonance contributions to the cross sections.}
\label{tab:3}
\begin{ruledtabular}
\begin{tabular}{l | c | c | c | c | c } 
Vector Resonance     & $\Gamma_{total}$ & BF($\mumu$)   &  Contribution to             & BF($\tautau$) &  Contribution to         \\
                     &   (MeV)          & (\%)          & $\sigma_{cuts}(\mumu)$ (\%)  & (\%)          &  $\sigma(\tautau)$ (\%) \\ \hline
$\Upsilon(4S)$(10580)&   20.5           &  0.0016       &   $<0.01$                    &  0.0016        &  $<0.01$                   \\
$\Upsilon(3S)$(10355)&   0.020          &  2.18         &   0.07                      &  2.29         &   0.07                  \\
$\Upsilon(2S)$(10023)&   0.032          &  1.93         &   0.04                      &  2.00          &   0.03                  \\
$\Upsilon(1S)$(9460) &   0.054          &  2.48         &   0.05                      &  2.60         &   0.05                  \\


$\Psi(2S)$(3686)     &    0.327         &  0.74         &   $<0.01$                       &  0.30         &   $<0.01$                \\
$J/\Psi(1S)$(3097)   &    0.093         &  5.93         &   0.12                       &  ...          &   ...                    \\ \hline
 Total               &                  &               &   0.28                       &               &   0.16                   \\ 
\end{tabular}
\end{ruledtabular}
\end{table*}

From this study it is evident that the vector meson resonances contribute at the 0.16\%  to the $\tau$-pair cross section
and the 0.28\% level to  $\sigma_{cuts}(\mumu)$. We assign these as conservative uncertainties on the \kk cross section calculation
because they are not included in the \kk code. Note that if an  $s^{\prime}$ cut is applied to a 
$\mu$-pair selection that removes the $J/\Psi(1S)$ and $\Psi(2S)$ resonances, 
e.g. $\rootsp > 0.4 \roots$, then this
component of the error on $\sigma_{cuts}(\mumu)$ is reduced to 0.15\%,
otherwise there will be an additional contribution of 0.12\% from $J/\Psi(1S)$ contribution to $\sigma_{cuts}(\mumu)$. 
If no cuts are applied, the error on $\sigma(\mumu)$ from this source is 0.33\%.

\section{Uncertainty on the ratio $\sigma(\tautau)$/$\sigma_{cuts}(\mumu)$}
\label{Ratio}

One of the means of determining the integrated luminosity of a data set at the B-factories is to count $\mu$-pair events.
If this luminosity is used to establish the number of produced $\tau$-pair events in the sample, then the ratio 
$\sigma(\tautau)$/$\sigma_{cuts}(\mumu)$ appears in the denominator of the branching fraction calculation. 
In this case, the error on this ratio contributes a theoretical systematic error associated with the overall
 normalization of the data sample. One can expect that some of the errors discussed in this paper
will  cancel in the ratio.

For the vacuum polarization uncertainty, the only component of uncertainty that survives in the ratio is associated
with the contributions to $\sigma(\mumu)$ below the \tautau threshold because
 above the \tautau threshold the vacuum polarization contributes to both \mtau-pairs and \mmu-pairs in the same way,
  apart from small effects related to the exact description of the $s^{\prime}$ dependence above the threshold.
We apply the same technique for evaluating the systematic error as
 described in Section~\ref{VacPol} except we only consider the contributions up to the \tautau threshold.
This yields an uncertainty of less than 0.05\%.
The uncertainty of the determination of the hadronic vacuum polarization
from the dispersion integral varies between 0.01\% and 0.02\% below
the \tautau threshold. These two uncertainties are independent and we
assign the total uncertainty coming from the vacuum polarization in the ratio 
$\sigma(\tautau)$/$\sigma_{cuts}(\mumu)$ to be 0.05\%.

For the treatment of initial and final state bremsstrahlung, we conservatively take as the error on the ratio the same value of 0.2\% 
assigned for the uncertainty on the absolute cross section. Almost certainly there are cancellations that further reduce 
this and which can be determined with additional study. Similarly, we assume the interference error on the ratio to be 0.04\%.
However, the pair-production and vertex correction uncertainties are 100\% correlated and largely cancel in the ratio.
Assuming the $\mu$-pair selection employed for a luminosity determination 
applies a cut of $\rootsp > 0.4 \roots$, the error associated with the resonances
will also cancel in the ratio, otherwise there will be an additional error of 0.12\%.

The total error we assign to the ratio $\sigma(\tautau)$/$\sigma_{cuts}(\mumu)=1.1000$ is 
$(0.05 \oplus 0.20 \oplus 0.04 \oplus 0.12 )\% = 0.24\%$.

\section{Summary}
\label{Summary}

Adding the uncertainties from the effects discussed in this paper in quadrature
we conclude that \kk calculates the \mtau-pair  and \mmu-pair
production cross sections at $\sqrt{s}=10.58~\gev$ with a
$(0.18 \oplus 0.20 \oplus 0.04 \oplus 0.15 \oplus 0.16)\% = 0.35\%$  and $(0.22 \oplus
0.20 \oplus 0.04 \oplus 0.15 \oplus 0.28)\% = 0.44\%$ relative calculational uncertainty, respectively.
 Using the IHVP flag = 4 option with the \kk program as
the central values, we obtain: $\sigma (\ee\to\tautau) = (0.919 \pm 0.003)\nb$,  $\sigma(\ee\to\mumu) = (1.147 \pm 0.005) \nb$,
 $\sigma_{cuts}(\ee\to\mumu) = (0.835 \pm 0.004) \nb$
and $\sigma(\tautau)$/$\sigma_{cuts}(\mumu)=1.100\pm0.003$.
Given the approach taken here to estimate the cross-section error,  it is evident that, if required,
 the quoted precision can be further improved in the future by undertaking a more detailed,
 though well-defined, analysis.

\vspace{12pt}
\centerline{\bf Acknowledgements}
 This work is supported in part  by  the Polish-French collaboration within IN2P3 through LAPP Annecy,  
 the EU 6$^{th}$ Framework Programme under contract MRTN-CT-2006-035482 (FLAVIAnet network),
 and the Natural Sciences and Engineering Research Council of
 Canada. We would like to thank H.~Burkhardt for the discussions
 on the implementation of vacuum polarization.



\begin{thebibliography}{99}

\bibitem{Hayasaka:2007vc}
  K.~Hayasaka {\it et al.}  [BELLE collaboration],
  arXiv:0705.0650 [hep-ex].

\bibitem{Yusa:2006qq}
  Y.~Yusa {\it et al.}  [BELLE Collaboration],
  Phys.\ Lett.\  B {\bf 640}, 138 (2006)
  [arXiv:hep-ex/0603036].


\bibitem{Jadach:1999vf}
  S.~Jadach, B.~F.~L.~Ward and Z.~Was,
  Comput.\ Phys.\ Commun.\  {\bf 130}, 260 (2000)
  [arXiv:hep-ph/9912214].

\bibitem{Jadach:1994ps}
  S.~Jadach and Z.~Was,
  Comput.\ Phys.\ Commun.\  {\bf 85}, 453 (1995).

\bibitem{Aubert:2005wa}
  B.~Aubert {\it et al.}  [\babar\ Collaboration],
  Phys.\ Rev.\ Lett.\  {\bf 96}, 041801 (2006)
  [arXiv:hep-ex/0508012].


\bibitem{Epifanov:2007rf}
  D.~Epifanov {\it et al.}  [Belle Collaboration],
  Phys.\ Lett.\  B {\bf 654}, 65 (2007)
  [arXiv:0706.2231 [hep-ex]].

\bibitem{Jezabek:1991qp}
 M.~Jezabek, Z.~Was, S.~Jadach and J.~H.~Kuhn,
 Comput.\ Phys.\ Commun.\  {\bf 70} (1992) 69.

\bibitem{Jadach:2000ir}
  S.~Jadach, B.~F.~L.~Ward and Z.~Was,
  Phys.\ Rev.\  D {\bf 63},  113009 (2001)
  [arXiv:hep-ph/0006359].


\bibitem{Ward:2002qq}
  B.~F.~L.~Ward, S.~Jadach and Z.~Was,
  Nucl.\ Phys.\ Proc.\ Suppl.\  {\bf 116}, 73 (2003)
  [arXiv:hep-ph/0211132].



\bibitem{Jadach:1991ws}
  S.~Jadach, B.~F.~L.~Ward and Z.~Was,
  Comput.\ Phys.\ Commun.\  {\bf 66}, 276 (1991).

\bibitem{Bardin:1989tq}
 D.~Y.~Bardin, M.~S.~Bilenky, T.~Riemann, M.~Sachwitz and H.~Vogt,
 Comput.\ Phys.\ Commun.\  {\bf 59} (1990) 303.

\bibitem{Bardin:1999yd}
 D.~Y.~Bardin, P.~Christova, M.~Jack, L.~Kalinovskaya, A.~Olchevski,
S.~Riemann and T.~Riemann,
 Comput.\ Phys.\ Commun.\  {\bf 133} (2001) 229
 [arXiv:hep-ph/9908433].

\bibitem{Kobel:2000aw}
  M.~Kobel {\it et al.}  [Two Fermion Working Group],
  [arXiv:hep-ph/0007180].

\bibitem{Eidelman:1995ny}
  S.~Eidelman and F.~Jegerlehner,
  Z.\ Phys.\  C {\bf 67}, 585 (1995)
  [arXiv:hep-ph/9502298].

\bibitem{Jegerlehner:1991ed}
  F.~Jegerlehner,
  Prog.\ Part.\ Nucl.\ Phys.\  {\bf 27}, 1 (1991).

\bibitem{Burkhardt:1989ky}
  H.~Burkhardt, F.~Jegerlehner, G.~Penso and C.~Verzegnassi,
  Z.\ Phys.\  C {\bf 43}, 497 (1989).

\bibitem{CrystalBall}
  C.~Edwards et al. [Crystal Ball Collaboration],
  SLAC-PUB-5160,  Jan. 1990.

\bibitem{Bai:2001ct}
  J.~Z.~Bai {\it et al.}  [BES Collaboration],
  Phys.\ Rev.\ Lett.\  {\bf 88}, 101802 (2002)
  [arXiv:hep-ex/0102003].

\bibitem{BuPie}
  H.~Burkhardt and B.~Pietrzyk,
  Phys.\ Lett.\  B {\bf 356}, 398 (1995);
   H.~Burkhardt and B.~Pietrzyk,
  Phys.\ Lett.\  B {\bf 513}, 46 (2001).

\bibitem{Burkhardt:2005se}
  H.~Burkhardt and B.~Pietrzyk,
  Phys.\ Rev.\  D {\bf 72}, 057501 (2005)
  [arXiv:hep-ph/0506323]; 
  REPI routine (version 2005):
  http://hbu.home.cern.ch/hbu/aqed/repi2005.f.

\bibitem{Karlen:2001hw}
  D.~Karlen and H.~Burkhardt,
  Eur.\ Phys.\ J.\  C {\bf 22}, 39 (2001)
  [arXiv:hep-ex/0105065].


\bibitem{Aubert:2004pwa}
  B.~Aubert {\it et al.}  [\babar\ Collaboration],
  Phys.\ Rev.\  D {\bf 72}, 032005 (2005)
  [arXiv:hep-ex/0405025].


\bibitem{KEKB}
Page 256 of Ref.~\cite{Yao:2006px} and Kenji Inami from Belle Collaboration (private communication).


\bibitem{Was:2006my}
  Z.~Was,
  arXiv:hep-ph/0610386.

\bibitem{Berends:1982ie}
  F.~A.~Berends, R.~Kleiss and S.~Jadach,
  Nucl.\ Phys.\  B {\bf 202}, 63 (1982).


\bibitem{Jadach:2001mp}
  S.~Jadach, W.~Placzek, M.~Skrzypek, B.~F.~L.~Ward and Z.~Was,
  Comput.\ Phys.\ Commun.\  {\bf 140}, 475 (2001)
  [arXiv:hep-ph/0104049].

\bibitem{Kuraev:1985hb}
  E.~A.~Kuraev and V.~S.~Fadin,
  Sov.\ J.\ Nucl.\ Phys.\  {\bf 41}, 466 (1985)
  [Yad.\ Fiz.\  {\bf 41}, 733 (1985)].

\bibitem{Benayoun:1999hm}
  M.~Benayoun, S.~I.~Eidelman, V.~N.~Ivanchenko and Z.~K.~Silagadze,
  Mod.\ Phys.\ Lett.\  A {\bf 14}, 2605 (1999)
  [arXiv:hep-ph/9910523].



\bibitem{Yao:2006px}
  W.~M.~Yao {\it et al.}  [Particle Data Group],
  J.\ Phys.\ G {\bf 33},  1 (2006)
  and 2007 partial update.



\bibitem{Aubert:2003sv}
  B.~Aubert {\it et al.}  [\babar\ Collaboration],
  Phys.\ Rev.\  D {\bf 69}, 011103 (2004)
  [arXiv:hep-ex/0310027].






\end{thebibliography}
\end{document}